# Observation of two-photon absorption at low power levels using tapered optical fibers in rubidium vapor

S.M. Hendrickson, M.M. Lai, T.B. Pittman and J.D. Franson
*Physics Department, University of Maryland, Baltimore County, Baltimore, MD 21250*

Nonlinear optical effects can be enhanced in tapered optical fibers with diameters less than the wavelength of the propagating light. Here we report on the observation of two-photon absorption using tapered fibers in rubidium vapor at power levels of less than 150 nW. Transit-time broadening produces two-photon absorption spectra with sharp peaks that are very different from conventional line shapes.

Tapered optical fibers with small diameters can produce relatively high intensities at low incident power levels due to their small mode area. This has allowed a number of recent demonstrations of nonlinear optical effects at low power levels [1-4]. Here we describe an experiment in which enhanced two-photon absorption [5,6] was observed in tapered optical fibers in the presence of rubidium vapor. Two-photon absorption was observed at power levels of less than 150 nW. To put this in perspective, these power levels correspond to less than 20 photons on average in the tapered region of the fiber at any given time.

Aside from its fundamental interest, two-photon absorption at low intensities may be useful for all-optical switching [2,7-10] or for quantum logic gates based on the Zeno effect [11-13]. The tapered optical fibers used in these experiments had diameters of 350 nm, which is less than half the wavelength of the light propagating in the fiber. As a result, atoms moving at thermal velocities pass through the evanescent field of the tapered region in a few nanoseconds, which produces interesting features in the shape of the two-photon absorption lines due to transit-time broadening.

Two-photon absorption in a three-level atom is illustrated schematically in Fig. 1. A photon at frequency $\omega_1$ is detuned from the resonant frequency of the first atomic transition by an amount $\delta$, which produces a virtual population of the second atomic state. A second photon at frequency $\omega_2$ gives a detuning $\Delta$ of the sum of the photon energies from the energy of the upper atomic state. We typically held $\omega_1$ and thus $\delta$ constant while scanning the frequency $\omega_2$ of the second photon. This produced a peak in the two-photon absorption rate when $\Delta = 0$. Unlike most previous experiments, we observed the two-photon absorption directly as a decrease in the transmission of the light at frequency $\omega_2$, rather than by monitoring the rate of fluorescent emission by the atoms.

Off-resonant two-photon absorption corresponds to the case in which $\delta$ is larger than the width of the intermediate atomic state, while resonant two-photon absorption [14-16] corresponds to $\delta = 0$. The rate of off-resonant two-photon absorption is proportional to $1/\delta^2$ for fixed power levels [5] and is thus considerably smaller than resonant two-photon absorption. We observed both types of two-photon absorption although the off-resonant two-photon absorption required somewhat higher power levels in order to maintain the signal to noise ratio.

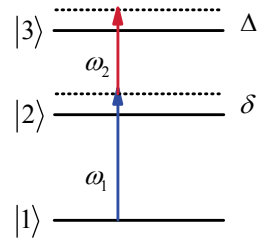

Fig. 1. Two-photon absorption in a three-level atom. We typically held the detuning $\delta$ of the intermediate state constant while scanning the detuning $\Delta$ of the final state.

Tapered optical fibers were fabricated from standard single-mode fiber using the well-known flame brush technique [17]. Our setup was designed to consistently fabricate tapered fibers with diameters of approximately 350 nm over a length of 5 mm. This diameter was chosen to coincide with the optimal combination of mode compression and evanescent power [5,18] for nonlinear interactions using the $5S_{1/2}$ to $5P_{3/2}$ to $5D_{5/2}$ transitions in rubidium, which have wavelengths of 780 and 776 nm, respectively.

To heat and pull the fibers to the desired diameter, we used an air and propane flame with a ¼" diameter nozzle with three 1 mm holes in a line perpendicular to the fiber axis. The air and fuel flow were stabilized using regulators in combination with digital mass flow controllers. The flame was oscillated by a computer controlled stage with a linear velocity profile. Transmission was monitored during the pulling process and we typically achieved 60% to 75% final transmission. The diameters of selected tapered fibers were measured between trials in a scanning electron microscope and they exhibited good agreement with the desired taper profiles.

After pulling, the fibers were mounted on stainless steel blocks with UV-curable epoxy suitable for high vacuum use. The blocks were attached to rails for structural support and to allow adjustment of the separation between the blocks. The mount was then attached to a flange and secured in a high vacuum system. Optical fibers passed through the vacuum flange using Teflon feedthroughs [19] and were fusion spliced to FC connectors for integration into an optical setup, as described below. The cumulative effect of the time spent in the atmosphere during mounting typically produced an additional 20% drop in total transmission, which has recently been attributed to the influence of ambient water vapor [20].

The vacuum system was pumped using a turbomolecular pump with an integrated scroll pump. The background pressure after bakeout was typically $10^{-8}$ Torr. To isolate the pump from the corrosive effects of rubidium we developed a procedure in which the system was baked-out and then sealed before rubidium was released into the chamber. A copper rod attached to a radiant heating element was positioned above the taper waist to reduce the accumulation of rubidium on the surface of the fiber. The primary chamber was kept near 100 ºC and the rubidium source at 125 ºC for the experiments reported here

The optical setup used for these experiments is shown in Fig. 2. Two frequency-stabilized external cavity diode lasers were tuned near the resonant frequencies of the $5S_{1/2}$ to $5P_{3/2}$ to $5D_{5/2}$ transition. The laser beams passed through optical isolators and then into single-mode fibers. The laser sources then passed through 50/50 fiber couplers to allow them to be used in a separate rubidium reference cell containing a double-pass configuration with transmission and fluorescence detection capabilities as well as a wavelength meter. The other ports of the fiber couplers passed through neutral density filters to attenuate the beams and then into the tapered fiber where they were counter-propagating.

we used the incident power (measured at points A and B in Fig. 2) as a conservative upper bound. The output power of the 776 nm beam was measured after it emerged from the vacuum chamber using a low-noise solid-state detector and a 1 nm filter to reject any reflected 780 nm signal. All unused ports on the couplers were fitted with fiber terminators to further reduce the effects of back-reflections. When necessary, the data acquisition system could average multiple data sets to reduce the noise. The system was also used to observe co-propagating two-photon absorption with minor changes to the setup.

Once a tapered optical fiber had been fabricated and inserted into the vacuum, we first evaluated the density of the surrounding rubidium vapor using a free-space beam that passed through the vacuum chamber via two glass windows. The free-space beam had a wavelength of 780 nm and it was used to measure the single-photon absorption due to transitions between the ground state and first excited state of the rubidium atoms. Similar measurements were then made using the transmitted power of a weak (<5 nW) field propagating in the tapered optical fiber itself. A typical scan of the single-photon absorption of the $D_2$ line of rubidium as measured by the transmission through the tapered fiber is shown in Fig. 3. The four dips correspond to the well-known ground-state hyperfine doublets of $^{85}$Rb and $^{87}$Rb [21]. Hyperfine splitting in the upper atomic levels was too small to be resolved in this experiment.

Spillane et al [1] recently demonstrated that the intensity enhancement due to the small mode area of a tapered optical fiber can result in the onset of saturation of the Rb $D_2$ line at low power levels. We also observed the onset of saturation at low power levels of typically 30 nW, where the features seen in Fig. 3 were reduced. The observed saturation power was consistent with that expected from the evanescent field of tapered fibers with these diameters.

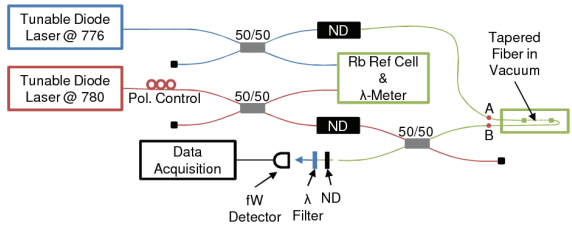

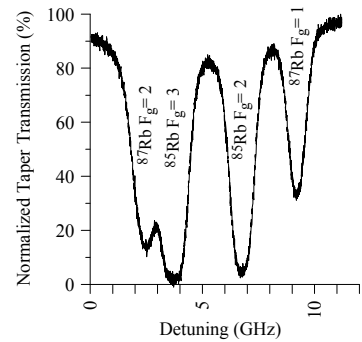

Fig. 2. Optical setup used to observe Doppler-free two-photon absorption in a tapered fiber. Counter-propagating beams passed through feedthroughs into a vacuum chamber containing a tapered fiber and rubidium atomic vapor and then back out again. 50/50 couplers allowed the beams to pass through a free-space reference cell for comparison purposes.

Fig. 3. A typical single-photon absorption scan used to verify the evanescent interaction between the guided mode in a tapered optical fiber and the surrounding Rb atoms. The transmission through a tapered optical fiber is plotted as a function of the detuning from the first excited state. The spacing of the dips is in agreement with the hyperfine splitting of the $5S_{1/2}$ ground state.

The intensity of the field in the tapered region was somewhat uncertain due to the transmission losses and

For comparison purposes, the off-resonant two-photon absorption spectrum obtained from fluorescence in the reference rubidium cell is shown in Fig. 4. The reference cell configuration used a mirror to reflect both free-space beams, so that there were equal components of co-propagating and counter-propagating photons. The sharp peaks in Fig. 4 correspond to the Doppler-free counter-propagating components while the broader peaks reflect the co-propagating Doppler broadening.

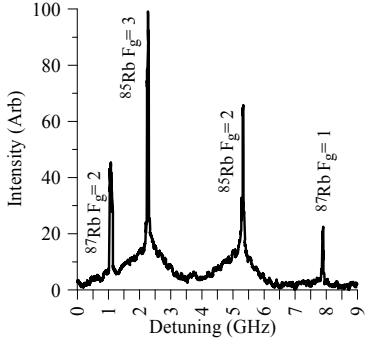

Fig. 4. Doppler-free two-photon absorption in a free-space rubidium reference cell with an off-resonant intermediate state. The fluorescence intensity from the decay of the upper atomic state is plotted as a function of the detuning (arbitrary offset). The broader peaks show the Doppler broadening for the co-propagating beams while the sharp peaks correspond to the Doppler-free spectrum for the counter-propagating beams.

Using the free-space rubidium cell as a frequency reference, we measured two-photon absorption using several tapered fibers. Fig. 5a shows an example of resonant [22, 23] two-photon absorption observed in a tapered optical fiber using counter-propagating beams. The transmission of the 776 nm signal was measured as a function of its detuning from the upper atomic level. Here the frequency of the 780 nm beam was held fixed on resonance with the $^{85}$Rb F=3 hyperfine state with an input power level of 146 nW. This power level was sufficiently low that it produced negligible power broadening, and the fitted width of the two-photon absorption dip was 97 MHz. The 776 nm input power was 29 nW. In Fig. 5b, the 780 nm input power level was increased to 726 nW, which increased the amount of two-photon absorption but also produced enough power broadening to increase the fitted linewidth to 122 MHz. This suggests that the power broadening was negligible in the data of Fig. 5a.

It can be seen from Fig. 5 that the dips in transmission due to two-photon absorption are very sharp and quite different from a Gaussian or a Lorentzian. Even if there were no Doppler broadening and the 780 nm field were exactly on resonance with an atom, the atom will pass through the region of the evanescent field in a few nanoseconds. This limits the time over which the atom interacts with the field and it produces a substantial amount of transit-time broadening, which has been characterized in recent work on hollow-core [24] and tapered optical fibers [1]. The expected two-photon absorption lineshape due to transit time broadening has the form [25]

$$\alpha(\Delta) \propto e^{-|\Delta \tau_0|} \quad (1)$$

Here $\Delta$ is once again the detuning from the upper atomic state and $\tau_0$ is the traversal time through the beam, which is approximately given by $\tau_0 = a/v_{th}$, where $a$ is the typical dimensions of the evanescent field region and $v_{th}$ is a typical value of the atomic thermal velocity. The data of Fig. 5a, where the power broadening was negligible, was fit using Eq. (1) with a value of $\tau_0 = 2.05$ ns.

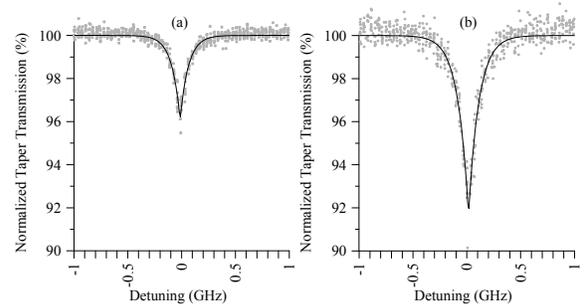

Fig. 5. Resonant two-photon absorption in a tapered optical fiber. The percent transmission of the 776 signal through the tapered fiber is plotted as a function of its detuning from the upper atomic state. (a) 780 nm power level of 146 nW. (b) 780 nm power level of 726 nW

Fig. 6a shows an example of Doppler-free two-photon absorption obtained with the 780 nm beam detuned off resonance from the $^{85}$Rb F=2 hyperfine ground state. The power in the tapered region was increased by approximately a factor of 20 in order to enhance the magnitude of the off-resonant two-photon absorption for the $^{85}$Rb F=3 hyperfine state. In addition to the large dip from the F=2 hyperfine line, a smaller off-resonant dip corresponding to the F=3 hyperfine state can be seen to the left of the main dip. Fig. 6b shows an example of results obtained using co-propagating photons, where it can be seen that the linewidth was broadened as compared to the counter-propagating case as expected [26]. As a result, the two dips can no longer be resolved and the line structure is asymmetric.

In order to put these power levels in perspective, we estimated the mean number of photons present in the tapered region at any given time. The energy in the taper can be calculated from the product of the propagation time through the tapered region (length 5 mm) and the total input power (~200 nW). The propagation time was calculated using the group velocity at 780 nm [27]. Combining these values gives an estimated energy in the tapered waist of $4.85 \times 10^{-18} J$ which corresponds to less than 20 photons at a time in the interaction region.

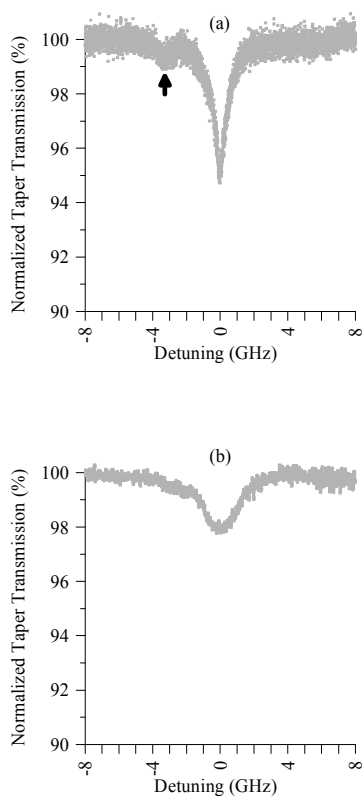

Fig. 6. Off-resonant two-photon absorption observed in a tapered optical fiber. (a) Counter-propagating two-photon absorption with high input power showing a near-resonant dip and an off-resonant dip (labeled by the arrow). (b) Co-propagating two-photon absorption showing increased linewidth and the inability to resolve the two lines.

In summary, we have demonstrated two-photon absorption in 350 nm diameter tapered optical fibers with incident power levels of less than 150 nW. The increase in the intensity due to the small mode area allows nonlinear effects of this kind at relatively low power levels. The sharp lineshape of the counter-propagating two-photon absorption dips is due to transit-time broadening. Two-photon absorption at low power levels may have practical applications in all-optical switching [28] or quantum logic gates [11-13].

We would like to acknowledge useful discussions with Alex Gaeta, Bryan Jacobs, Prem Kumar, and Selim Shahriar. This work was funded by the DARPA Defense Sciences Office.

1. S. M. Spillane et al., *Phys Rev Lett* **100**, 233602 (2008).
2. M. Bajcsy et al., *Phys. Rev. Lett.* **102**, 203902 May (2009).
3. F. Benabid, J. C. Knight, G. Antonopoulos, and P. St. J. Russel, *Science* **298**, 399-402 (2002).
4. W. Yang et al., *Nature Photon.* **1**, 331 (2007).
5. H. You, S. M. Hendrickson, and J. D. Franson, *Phys. Rev. A* **78**, 053803 (2008).
6. H. You, S. M. Hendrickson, and J. D. Franson, *Phys. Rev. A* **80**, 043823 (2009).
7. K. Salit, M. Salit, S. Krishnamurthy, R. Beausoleil, and M. S. Shahriar, *Preprint*.
8. D. A. Braje, V. Balic, G. Y. Yin, and S. E. Harris, *Phys. Rev. A* **68**, 041801(R) (2003).
9. A. M. Dawes, L. Illing, S. M. Clark, and D. J. Gauthier, *Science* **308**, 672-674 April (2005).
10. P. Londero, V. Venkataraman, A. R. Bhagwat, A. D. Slepkov, and A. L. Gaeta, *Phys. Rev. Lett.* **103**, 043602 July (2009).
11. J. D. Franson, B. C. Jacobs, and T. B. Pittman, *Phys. Rev. A* **70**, 062302 (2004).
12. J. D. Franson, B. C. Jacobs, and T. B. Pittman, *Phys. Rev. A* **70**, 062302 (2004).
13. J. D. Franson, B. C. Jacobs, and T. B. Pittman, *J. Opt. Soc. Am. B* **24**, 209-213 (2007).
14. R. Salomaa and S. Stenholm, *J. Phys. B: Atom. Molec. Phys.* **8**(11), 1795-1805 (1975).
15. R. Salomaa and S. Stenholm, *Applied Phys.* **17**, 309-316 (1978).
16. J. E. Bjorkholm and P. F. Liao, *Phys. Rev. Lett.* **33**, 128 (1974).
17. T. A. Birks and Y. W. Li, *J. Lightwave Technol.* **10**(4), 432-438 (1992).
18. M. A. Foster, A. C. Turner, M. Lipson, and A. L. Gaeta, *Opt. Expr.* **16**(2), 1300 (2008).
19. E. Abraham and E. Cornell, *Appl. Opt.* **37**, 1762-1763 (1998).
20. L. Ding, C. Belacel, S. Ducci, G. Leo, and I. Favero, *Applied Optics* **49**(13), 2441-2445 (2010).
21. P. Siddons, C. S. Adams, C. Ge, and I. G. Hughes, *J. Phys. B: At. Mol. Opt. Phys.* **41**, 155004 (2008).
22. T. T. Grove, V. Sanchez-Villicana, B. C. Duncan, S. Maleki, and P.L. Gould, *Physica Scripta.* **52**, 271-276 (1995).
23. P. R. Berman, *Phys. Rep.* **43**(3), 101-149 (1978).
24. A. D. Slepkov, A. R. Bhagwat, V. Venkataraman, P. Londero, and A. Gaeta, *Phys. Rev. A* **81**, 053825 (2010).
25. S. N. Bagayev, V. P. Chebotayev, and E. A. Titov, *Laser Physics* **4**(2), 224-292 (1994).
26. J. E. Bjorkholm and P. F. Liao, *Phys. Rev. A* **14**(2), 751-760 (1976).
27. L. Tong, J. Lou, and E. Mazur, *Opt. Express* **12**, 1025-1035 (2004).
28. B.C. Jacobs and J.D. Franson, , Phys. Rev. A **79**, 063830 (2009).